\documentclass[12pt]{article}

\def\tbgamma{\tilde{B}_{\tilde{\gamma}}}
\def\bgamma{B_{\gamma}}

\def\tgamma{\tilde{\gamma}}
\def\pp{M}

\usepackage{graphicx}
\input epsf

\begin{document}

\begin{titlepage}
\begin{flushright}
CALT-68-2568\\
ITEP-TH-50/05\\
\end{flushright}

\begin{center}
{\Large\bf $ $ \\ $ $ \\
B\"acklund transformations,  energy shift\\[5pt] 
and the plane wave limit.}\\
\bigskip\bigskip\bigskip
{\large Andrei Mikhailov}
\\
\bigskip\bigskip
{\it California Institute of Technology 452-48,
Pasadena CA 91125 \\
\bigskip
and\\
\bigskip
Institute for Theoretical and 
Experimental Physics, \\
117259, Bol. Cheremushkinskaya, 25, 
Moscow, Russia}\\

\vskip 1cm
\end{center}

\begin{abstract}
	We discuss basic properties of the B\"acklund transformations
	for the classical string in AdS space in the context of the
	null-surface perturbation theory.
	We explain the relation between
	the B\"acklund transformations and the energy shift
	of the dual field theory state.
	We show that the B\"acklund transformations
	can be represented as a finite-time evolution
	generated by a special linear combination
	of the Pohlmeyer charges. This is a manifestation
	of the general property of B\"acklund transformations
	known as spectrality. We also discuss the
	plane wave limit. 
\end{abstract}

\end{titlepage}

\section{Introduction}
One of the main goals of the AdS/CFT correspondence is to gain insight
in the dynamics of the string theory in backgrounds with nonzero Ramond-Ramond
field strength. Integrability is important in the AdS/CFT
program \cite{Minahan:2002ve,Mandal:2002fs,Bena:2003wd,Alday:2003zb}. 
Our understanding of classical integrability of the string worldsheet
is probably somewhat incomplete at this point, but even the current
results are already very impressive.  Many explicit solutions are known,
and in some sense we have the full construction of the action-angle
variables in the finite-gap approach 
\cite{Kazakov:2004qf,Kazakov:2004nh,Beisert:2004ag,Beisert:2005bm,%
Schafer-Nameki:2004ik}. 
There is a remarkable partial agreement with the Yang-Mills perturbative
calculations of the anomalous dimension \cite{FT02,Tseytlin,Russo,FT03}. 

The most important goal in 
the classical theory of integrability is to identify
the integrable structures which have a transparent meaning
in the quantum theory.
In the integrable string theory, we eventually want to 
be able to generalize the integrable structures from the sphere
to the higher genus surfaces. From this point of view, it would be
useful to understand the integrability as much as possible
in terms of the objects which are local on the worldsheet.
This would be also important if we want to compare the
string theory computations to the Yang-Mills computations,
because the Yang-Mills diagramms are local in a sense that
they involve only the interactions of those partons
which are close neighbors on the spin chain.

Examples of those objects which are local on the worldsheet
are local conserved charges and
B\"acklund equations. Local conserved charges were constructed
by Pohlmeyer, and in fact B\"acklund transformations were
used to define them \cite{Pohlmeyer:1975nb}
and to actually compute them\footnote{Local conserved charges
can be also obtained from the eigenvalues of the monodromy matrix;
but in practice the shortest way to write them explicitly
is probably to use their definition through B\"acklund
transformations.} for particular solutions \cite{Arutyunov:2003rg}. 
B\"acklund transformations
allow us to construct the new solution from a given solution,
as a solution of the  differential equation which we will call
the B\"acklund equation.  
The B\"acklund equation depends on the real parameter $\gamma$.
It is of the form
\begin{equation}\label{GeneralForm}
	\partial\phi_{new}=\Phi_{\gamma}(\phi_{new},\phi_{old},
	\partial\phi_{old}) 
\end{equation}
where $\phi$ are the embedding functions of the string worldsheet 
 into the target space and $\partial$ stands for the derivatives
 with respect to the worldsheet coordinates.
Solving the B\"acklund equation involves
the choice of the integration constants. It turns out that
for some particular value of the integration constants
$\phi_{new}$ is in fact a Hamiltonian flow of
$\phi_{old}$ by a certain infinite linear combination of
the local conserved charges.
The coefficients of this infinite linear combination depend on
$\gamma$.  For every $\gamma$ we have a Hamiltonian $H_{\gamma}$
and the corresponding Hamiltonian vector field $\xi_{\gamma}$
such that the flow by the finite time
is the B\"acklund transformation.
This means that, even though the local conserved charges can appear
to be
complicated in form, in fact the Hamiltonian flows generated
by certain combinations of these charges by a finite time are controlled by 
the explicitly known differential equation of the form 
(\ref{GeneralForm}) which is in fact closely related to the 
auxiliary linear problem\footnote{
 The trick is known
in matrix models as passing from the ordinary times $t_n$ to ``Miwa times''
$\gamma_p$: 
$t_n=\sum_p {\gamma_p^n\over n}$. Introducing $\gamma_p$
corresponds to creation of the fermion $\psi(\gamma_p)$
from the Fermi sea. This approach was developed for example in 
\cite{Kharchev:1991cy}.}.

B\"acklund transformations are important in the quantum theory 
\cite{Gaudin:1992ci,SklyaninBQ,Bazhanov:1996dr}.
The Hamiltonian is usually related to the
quantum B\"acklund transformation  by the Baxter's
$T-Q$ relation.
In the context of AdS/CFT correspondence the natural object
is not the Hamiltonian (which would not be conformally invariant)
but the discrete\footnote{Integrability of discrete canonical
transformations was discussed in \cite{Veselov}.}
``deck transformation'' which corresponds
to the anomalous dimension on the Yang-Mills side
(see the discussion in Section \ref{sec:DeckGeneral} and
in \cite{Mikhailov:2004ca}).
As we will see in the Section \ref{sec:Lift}, the deck transformation
is literally a particular example of a B\"acklund transformation.
It would be very interesting to see if there is an analogue
of this fact for the Yang-Mills diagramms\footnote{The role
of the local conserved charges in the Yang-Mills computations
was discussed in \cite{Dolan:2003uh,Dolan:2004ps,Dolan:2004ys}.}.

In Section \ref{sec:BA}  we discuss
the basic properties of B\"acklund transformations,
prove their canonicity and explain the relation between
B\"acklund transformations and deck transformations.
We also argue that a B\"acklund transformation 
can be represented as a Hamiltonian flow by a finite time
generated by an infinite linear combination of the local conserved
charges. (We will call this infinite linear combination
the ``generator'' of the B\"acklund transformation.)
Some of the results of Section \ref{sec:BA}
have already been presented in \cite{Mikhailov:2004ca,Mikhailov:2005qv}. 
In Section \ref{sec:BP} we derive the explicit formula 
(\ref{Spectrality}), (\ref{ExplicitFormula})
for the generator.
We present two methods of deriving this formula: using the plane
wave limit and using the formula (\ref{DeltaOfSomething}) for 
the variation of the 
symplectic potential under the B\"acklund transformation.
We also use the results
of \cite{Arutyunov:2005nk}  to
express the formula (\ref{IntegralFormulaForHInfty})
for the anomalous dimension 
conjectured in \cite{Mikhailov:2005wn} in terms of 
the eigenvalues of the monodromy matrix.
We use the properties of B\"acklund transformation to prove
this conjectured formula.

We use the plane wave limit of the local conserved charges
and B\"acklund transformations. This limit has a clear physical meaning
explained in \cite{BMN}.
In this limit the string worldsheet theory becomes the
theory of free massive fields. Although this is just a limit,
in some sense it ``captures'' all the conserved charges, at least the
local charges.
Therefore we can use the plane wave limit to guess, and even prove,
various relations between the conserved charges which hold beyond this limit.

\section{Anomalous dimension as a B\"acklund transformation.}
\label{sec:BA}
\subsection{Deviation of the string worldsheet from being periodic.}
An important property of the AdS space is its ``periodicity''.
It is often convenient to introduce the global coordinates in which
the metric becomes:
$$
	ds^2=-(1+r^2)dt^2+{dr^2\over 1+r^2} + r^2 d\Omega_{S^{n-1}}^2
$$
In these coordinates the periodicity corresponds to the
discrete symmetry $t\to t+2\pi$. 
It is important that this discrete symmetry commutes
with all the global symmetries and in fact belongs to the
center of $SO(2,n)$.
We will call it ``deck transformation'', because it is a deck transformation
if we think of $AdS_{n+1}$ as the universal covering space of the hyperboloid.

Consider the Type IIB superstring theory on $AdS_5\times S^5$.
It is useful to look at the deviation of the dynamics of the theory
from being periodic in $t$. For example, solutions of the linearized
classical supergravity equations are exactly periodic 
(invariant under $t\mapsto t+2\pi$), although the solutions of the
nonlinear classical supergravity are generally speaking not periodic.
Consider a classical string moving in $AdS_5\times S^5$.
If we neglect the backreaction of the string on the AdS geometry,
then it is possible to quantitatively characterize the deviation
of the string worldsheet from being periodic in $t$.

Indeed, the transformation $t\mapsto t+2\pi$ is a discrete symmetry
and we can ask ourselves how it acts on the string phase space.
The classical superstring on $AdS_5\times S^5$ is an integrable
model, there are infinitely many conserved charges in involution.
In some sense these charges generate the ``invariant tori''.
This is precisely true for the so-called finite-gap solutions.
The Hamiltonian flows of these finite-gap solutions fill
the finite-dimensional tori. (For the general solutions we can 
probably talk about the ``infinite-dimensional tori'' but it is not very
clear what these words mean.) 
If we represent the invariant torus as a quotient
$$
	{\bf T}^g={\bf R}^g/({\bf Z} e_1 +\ldots +{\bf Z} e_g)
$$
then the transformation $t\mapsto t+2\pi$ would act as a shift
by the vector $a_1 e_1+\ldots +a_g e_g$. The numbers $a_1,\ldots, a_g$
characterize quantitatively the deviation of the string
worldsheet from being periodic in $t$.

The set of points ${\bf Z}(a_1 e_1+\ldots +a_g e_g)$ is usually
a dense subset of the invariant torus, and therefore it actually
{\em defines} the torus (the torus is the ``closure'' of this set).
This suggests that the action of this discrete symmetry carries
a lot of information about the dynamics of the string.
On the field theory side this symmetry measures the
anomalous dimension of the corresponding CFT operator. 

\subsection{B\"acklund transformations.}
\subsubsection{Definition of B\"acklund transformations.}
\label{sec:Definition}
The classical string is the nonlinear sigma-model with the Virasoro
constrains. Most of the discussion in our paper is valid
independently of the Virasoro constraints.
If we turn off the fermions, the nonlinear sigma-model on 
$AdS_5\times S^5$
is essentially the product of the sigma-models on $AdS_5$ and
$S^5$. We will concentrate on the $AdS_5$-part
of the sigma-model.
Let us parametrize the points on $AdS_5/{\bf Z}$ by
the unit vectors $X^{\mu}$, $\mu\in {-1,0,1,2,3,4}$, subject
to the constraint $g_{\mu\nu}X^{\mu}X^{\nu}=1$ with
$g_{\mu\nu}=\mbox{diag}(1,1,-1,-1,-1,-1)$. This means that we
realize
$AdS_5/{\bf Z}$ as the hyperboloid in ${\bf R}^{2+4}$.
In this subsection we will consider the B\"acklund transformations
acting on the projection of
the string worldsheet to the hyperboloid $AdS_5/{\bf Z}$.
We will consider the lift of the B\"acklund transformations
to the string on  $AdS_5$ in Section \ref{sec:Lift}.

There are infinitely many Pohlmeyer charges ${\cal F}_{2n}$
and $\widetilde{\cal F}_{2n}$ introduced in $\cite{Pohlmeyer:1975nb}$
and the corresponding Hamiltonian vector fields 
will be denoted $\xi_{ {\cal F}_{2n} }$
and $\xi_{\widetilde{\cal F}_{2n} }$.
We will give a definition of the Pohlmeyer charges 
in Section \ref{sec:DefinitionOfCharges}, but here we just
point out that they are the local conserved charges.
A local conserved charge is given by an integral over the closed 
contour on the string worldsheet of a closed differential
1-form, which is constructed from the worldsheet
fields $X(\tau, \sigma)$ and their derivatives.
The integral does not depend on the choice of the contour
because the 1-form is closed. In this sense, it is a conserved charge.
The existence of infinitely many closed differential 1-forms
on the worldsheet is a very nontrivial fact related to the
integrability of the worldsheet theory.

Given a vector field $\xi$ on the phase space we can consider the
flow of this vector field. It is a one-parameter family
of transformations which we will denote:
\[e^{s\xi}\;\;\mbox{or}\;\;\exp(s\xi)\]
Given a point $x$ of the phase space, the
flow by the time $s$ of this point, denoted
$e^{s\xi}x$, is defined as $y(s)$ where $y(t)$ is the solution of
the differential equation ${dy(t)\over dt}=\xi(y(t))$ with
the initial condition $y(0)=x$.

We introduce the conformal coordinates $\tau$ and $\sigma$
on the worldsheet so that the metric is proportional
to $d\tau^2- d\sigma^2$. We denote 
$\partial_{\pm}={1\over 2}(\partial_{\tau}\pm\partial_{\sigma})$.
Given the string worldsheet $X^{\mu}(\tau,\sigma)$
we consider two discrete transformations $B_{\gamma}X$
and $\tilde{B}_{\tilde{\gamma}}X$ depending on 
real parameters $\gamma$ and $\tilde{\gamma}$,
which are defined by the following
properties:
\begin{enumerate}
	\item These transformations are generated by 
		the Hamiltonian flows\footnote{The interpretation
		of B\"acklund transformations as shifts
		on the Jacobian was discussed in 
		\cite{KuznetsovVanhaecke}.} of the Pohlmeyer charges:
	\begin{eqnarray}
		B_{\gamma}X=
		\exp\left(\sum_n t_{2n}({\gamma}) 
		\xi_{ {\cal F}_{2n} }\right).X
		\label{LeftContinuous}
		\\[5pt]
		\tilde{B}_{\tilde{\gamma}}X=
		\exp\left(\sum_n \tilde{t}_{2n}(\tilde{\gamma})
		\xi_{\widetilde{\cal F}_{2n}}\right).X
		\label{RightContinuous}
	\end{eqnarray}
	with some coefficients $t_{2n}$ and $\tilde{t}_{2n}$
	depending on $\gamma$ and $\tilde{\gamma}$.
\item They satisfy the following first order differential equations:
\begin{equation}\label{LBX}
	\begin{array}{l}
		\partial_-(B_{\gamma}X-X)=
	-{1\over 2}(1+\gamma^2)(B_{\gamma}X,\partial_-X)
	(B_{\gamma}X+X)
	\\[5pt]
	\partial_+(B_{\gamma}X+X)=
	{1\over 2}(1+\gamma^{-2})(B_{\gamma}X,\partial_+X)
	(B_{\gamma}X-X)
\end{array}
\end{equation}
\begin{equation}\label{RBX}
	\begin{array}{l}
\partial_+(\tbgamma X -X)=
-{1\over 2}(1+\tgamma^{-2})(\tbgamma X,\partial_+X)
(\tbgamma X+X)
	\\[5pt]
	\partial_{-}(\tbgamma X+X)=
	{1\over 2}(1+\tgamma^2)(\tbgamma X,\partial_-X)
	(\tbgamma X-X)
\end{array}
\end{equation}
\item For small $\gamma$ and large $\tgamma$: 
\begin{eqnarray}
&&	\bgamma X=X-\gamma {\partial_+X\over |\partial_+X|}+
o(\gamma)\label{SmallGamma}\\[5pt]
&&	\tbgamma X=X+{1\over\tgamma}{\partial_-X\over |\partial_-X|}
	+o\left({1\over\tgamma}\right)
	\label{LargeGamma}
\end{eqnarray}
\end{enumerate}
We conjecture that the transformations
$\bgamma$ and $\tbgamma$ exist and are determined unambiguously by 
these three properties, at least if the velocity of the
string is large enough. 
The coefficients $t_n(\gamma)$ and $\tilde{t}_n(\tgamma)$ 
will be determined in Section \ref{sec:BP}. 
The relation between B\"acklund transformations
and the Hamiltonian vector fields generated by the local conserved charges 
in the special case when the motion of the
string is restricted to ${\bf R}\times S^2$ 
was discussed in \cite{Mikhailov:2005qv}. 

\subsubsection{Perturbative solutions of B\"acklund equations.}
\label{PerturbativeSolutionsOfBE}
Eqs. (\ref{LBX}), (\ref{RBX}), (\ref{SmallGamma}) and
(\ref{LargeGamma}) can be solved as the series in $\gamma$
and $1/\tgamma$ when $\gamma$ is small and $\tgamma$ is
large. The recurrent relations allowing to construct
the solution order by order in $\gamma$ and $1/\tgamma$
were written for example in \cite{Arutyunov:2003rg}.

Another way of solving (\ref{LBX}) and
(\ref{RBX}) perturbatively is to use the null-surface
perturbation theory 
\cite{DeVega:1992tm,deVega:1994hu,Mikhailov:2003gq,Kruczenski:2004cn,%
Mikhailov:2004au}
with the small
parameter $1/|\partial_{\tau}X|$. (More precisely, the small parameter
is the angle between $\partial_+X$ and $\partial_-X$; in the
null-surface limit $\partial_+X=\partial_-X$. If we define
the conformal coordinates $(\tau,\sigma)$ so that 
$|\partial_{\sigma}X|\simeq 1$ then
the small parameter is $1/|\partial_{\tau}X|$.
Alternatively, if we define the coordinates so that
$|\partial_{\tau}X|\simeq 1$, then the small parameter is
$|\partial_{\sigma}X|$.)
The zeroth order is:
\begin{eqnarray}
&&	\bgamma X={1-\gamma^2\over 1+\gamma^2}X
	-{2\gamma\over 1+\gamma^2}{\partial_{\tau}X\over |\partial_{\tau}X|}
	+\ldots
	\label{LeftZerothOrder}
	\\[5pt]
&&	\tbgamma X={1-\tgamma^{-2}\over 1+\tgamma^{-2}}X
	+{2\tgamma^{-1}\over 1+\tgamma^{-2}}
	{\partial_{\tau}X\over |\partial_{\tau}X|}
	+\ldots
	\label{RightZerothOrder}
\end{eqnarray}
where dots denote the terms of the higher order in $1/|\partial_{\tau}X|$.
We believe that the series of the null-surface perturbation theory
converge if the string moves fast enough, but we do not have a proof
of that. 

The null-surface perturbation theory works for finite values of
$\gamma$ and $\tilde{\gamma}$. But if $\gamma$ is small, then
the null-surface perturbation theory agrees with the perturbative
expansion in $\gamma$, in the following sense. Any finite order
of the null-surface expansion is a rational function
of $\gamma$, and can be expanded as a Taylor series around $\gamma=0$.
These series in $\gamma$
converge in some finite region of the parameter $\gamma$
around $\gamma=0$. We will now discuss some details of the 
null-surface perturbation theory for $B_{\gamma}X$ and 
explain why any finite order is a rational
function of $\gamma$. 

Suppose that $1/M$ is a small parameter
of the null-surface perturbation theory, and choose the
coordinates $\tau$, $\sigma$ so that:
\[
|\partial_{\tau}X|\simeq 1,\;\;\; |\partial_{\sigma}X|\simeq 1/M
\]
We can introduce the new worldsheet coordinate $s=\sigma/M$ so that
$|\partial_s X|\simeq 1$ and 
$\partial_{\pm}={1\over 2}(\partial_{\tau}\pm M^{-1}\partial_s)$.
Subtracting the first equation of (\ref{LBX}) from the second
equation of (\ref{LBX}) we get
\begin{equation}\label{Iterative}
	B_{\gamma}X-{1-\gamma^2\over 1+\gamma^2}X
	-{4\partial_{\tau}X\over 
	(\gamma+\gamma^{-1})^2(B_{\gamma}X,\partial_{\tau}X)}
	=
	{
	M^{-1}\Phi[X,B_{\gamma}X]
	\over (\gamma+\gamma^{-1})^2(B_{\gamma}X,\partial_{\tau}X)}
\end{equation}
where
\[
\Phi[X,B_{\gamma}X]=4\partial_s B_{\gamma}X+
(\gamma^2-\gamma^{-2})(B_{\gamma}X,\partial_s X)B_{\gamma}X
+(\gamma+\gamma^{-1})^2(B_{\gamma}X,\partial_sX)X
\] 
The null-surface perturbation
theory can be understood as  solving Eq. (\ref{Iterative})
for $B_{\gamma}X$ by an iterative procedure, taking as the zeroth
order ($M=\infty$):  $B_{\gamma}^{(0)}X={1-\gamma^2\over 1+\gamma^2}X
-{2\gamma\over 1+\gamma^2}{\partial_{\tau}X\over |\partial_{\tau}X|}$.
Given $B_{\gamma}^{(n)}$, we define $B_{\gamma}^{(n+1)}$
by the formula
\[
B_{\gamma}^{(n+1)}X
-{1-\gamma^2\over 1+\gamma^2}X
-{4\partial_{\tau}X\over
(\gamma+\gamma^{-1})^2(B^{(n+1)}_{\gamma}X,\partial_{\tau}X)}
=
	{
	M^{-1}\Phi[X,B^{(n)}_{\gamma}X]
	\over (\gamma+\gamma^{-1})^2(B^{(n)}_{\gamma}X,\partial_{\tau}X)}
\]
This equation defines $B^{(n+1)}_{\gamma}X$ in terms of
$X$ and $B^{(n)}_{\gamma}X$ in the following way.
We represent 
$B^{(n+1)}_{\gamma}X=(B^{(n+1)}_{\gamma}X)_{\perp}+
S^{(n+1)}_{\gamma}{\partial_{\tau}X\over |\partial_{\tau}X|^2}$,
where  
\[
(B^{(n+1)}_{\gamma}X)_{\perp}={1-\gamma^2\over 1+\gamma^2}X
+\left[{M^{-1}\Phi[X,B^{(n)}_{\gamma}X]
	\over (\gamma+\gamma^{-1})^2(B^{(n)}_{\gamma}X,\partial_{\tau}X)}
	\right]_{\perp}
	\]
	the subindex $\perp$ meaning the component of the vector
	orthogonal to $\partial_{\tau}X$, and 
	$S^{(n+1)}_{\gamma}=(B^{(n+1)}_{\gamma}X,\partial_{\tau}X)$
	is determined from the quadratic equation:
\[
S^{(n+1)}_{\gamma}-
{4|\partial_{\tau}X|^2\over (\gamma+\gamma^{-1})^2 S^{(n+1)}_{\gamma}}
=	
{M^{-1}(\Phi[X,B^{(n)}_{\gamma}X],\partial_{\tau}X)
\over (\gamma+\gamma^{-1})^2(B^{(n)}_{\gamma}X,\partial_{\tau}X)}
\]
and we write the solution to this equation as a power series in
$M^{-1}$:
\[
S^{(n+1)}_{\gamma}=-{2\over \gamma+\gamma^{-1}}|\partial_{\tau}X|
+{1\over 2}{M^{-1}(\Phi[X,B^{(n)}_{\gamma}X],\partial_{\tau}X)
\over (\gamma+\gamma^{-1})^2(B^{(n)}_{\gamma}X,\partial_{\tau}X)}+
\ldots
\]
One can prove by induction in $n$
that  $(B^{(n)}_{\gamma}X,X)={1-\gamma^2\over 1+\gamma^2}$, and also that 
$B^{(n+1)}_{\gamma}X=B^{(n)}X+O(M^{-n-1})$.
It is important here that $\partial_s^pX$
and $\partial_s^p\partial_{\tau}X$ are of the order $1$
for any $p$ (derivatives of $X$ do not introduce positive
powers of $M$).
This is controlled by the initial conditions on $X$, and
this is implied when we say that the
string is fast moving. 

The null-surface perturbation theory calculates
$B_{\gamma}X$ as $B_{\gamma}^{(\infty)}X$.
Eq. (\ref{Iterative}) implies that $| B_{\gamma}^{(\infty)}X |^2=1$.
Indeed, taking the scalar product of (\ref{Iterative}) with
$B_{\gamma}X$ we get:
\begin{equation}\label{BGammaSquare}
	\begin{array}{c}
	1-|B_{\gamma}X|^2=
	{2M^{-1}\over (\gamma+\gamma^{-1})^2(B_{\gamma}X,\partial_{\tau}X)
	-M^{-1}
	(\gamma^2-\gamma^{-2})(B_{\gamma}X,\partial_s X)}
	\partial_s(1-|B_{\gamma}X|^2)
\end{array}
\end{equation}
Assuming that that $1-|B_{\gamma}X|^2$ is of the order $1/M^k$ leads
to an immediate contradiction with (\ref{BGammaSquare}) because
the right hand side would be then of the order at least as small as $1/M^{k+1}$.
This means that $|B_{\gamma}X|^2=1$ as a power series in $M^{-1}$.
 
Only the rational functions of $\gamma$ with the denominator
the power of $\gamma+\gamma^{-1}$ appear in this iterative procedure.
(The other thing we will have to divide by is the powers of 
$|\partial_{\tau}X|$.)
This implies that in the null-surface expansion $B_{\gamma}X$ is
a rational function of $\gamma$ with the poles
at $\gamma=\pm i$.

The iterative procedure uses only the difference of the first
and the second equations in (\ref{LBX}).
The sum of the first and the second equations will be satisfied
automatically for the perturbative solution, because 
of the consistency of the B\"acklund equations.

\subsubsection{General B\"acklund transformations and canonical
B\"acklund transformations.}
The definition of the B\"acklund transformation which we use here
is slightly different from the one usually accepted. 
Solving Eqs. (\ref{LBX}) involves the choice of the integration constants.
Usually any solution $B_{\gamma}X$ of (\ref{LBX}) is called
the B\"acklund transform of $X$. However most of the solutions of
the B\"acklund equations cannot be represented
by the flow  of $X$ by a linear
combination of the local charges as in (\ref{LeftContinuous}). From many points of view
the most general solutions of (\ref{LBX}) are more
interesting than the special solutions which we consider here,
because they allow to construct essentially new solutions from 
the known solution. (While the special solutions $B_{\gamma}X$ which 
we consider here are just the ``shift of times'' 
\cite{Mikhailov:2005qv} of $X$.)
The special solutions of (\ref{LBX}) corresponding to the
shift of times can be described in three ways:
\begin{enumerate}
	\item Solve (\ref{LBX}) perturbatively using $\gamma$
		as a small parameter and verify that the
		series converge.
	\item Solve (\ref{LBX}) in the null-surface perturbation
		theory and verify that the series converge.
	\item Impose periodic boundary conditions 
		$X(\tau,\sigma+2\pi)=X(\tau,\sigma)$ and 
		find the solution of (\ref{LBX}) which satisfies
		these boundary conditions.
\end{enumerate}
These three methods should give the same result, at least
if the string moves fast enough.
We are interested in these special or ``perturbative'' solutions
of the B\"acklund equations because they carry the information
about the Hamiltonian flows generated by the local conserved charges.

For these ``perturbative'' B\"acklund transformations  
Eqs. (\ref{LeftContinuous}), (\ref{RightContinuous}) actually
follow from the B\"acklund equations (\ref{LBX}), (\ref{RBX}).
Indeed, 
we will prove in Section \ref{sec:Canonicity} that for the ``perturbative''
B\"acklund transformations Eqs. (\ref{LBX}) and (\ref{RBX}) imply
the canonicity. 
(In fact, the proof uses the periodicity of $B_{\gamma}X$ and 
$\tilde{B}_{\tilde{\gamma}}X$.)
In Section \ref{sec:Logarithm} we explain how to
define the vector fields $\xi_{\gamma}$ and $\tilde{\xi}_{\tgamma}$
given the transformations $B_{\gamma}$ and $\tilde{B}_{\tilde{\gamma}}$
defined by (\ref{LBX}) and (\ref{RBX}). Because of the canonicity
of the B\"acklund transformations, 
these vector fields should be generated by an infinite linear combination
of the local conserved charges. In Sections \ref{sec:DefinitionOfCharges}
and \ref{sec:NoOtherCharges} we will describe these local conserved charges,
and in Section 
\ref{sec:Generator} fix the coefficients of this linear combination
(see Eq. (\ref{Spectrality})).
This gives us Eqs. (\ref{LeftContinuous}) and (\ref{RightContinuous}).
But nevertheless we have
decided to include Eqs. (\ref{LeftContinuous}) and (\ref{RightContinuous})
in the definition of B\"acklund transformations.

\begin{figure}
%	\begin{center}
\centering
\includegraphics[height=2.1in]{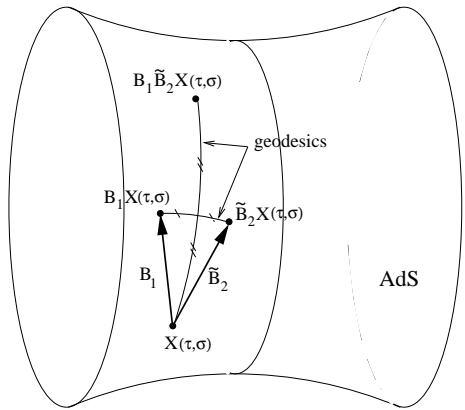}
\caption{The superposition of two B\"acklund transformations.
If we know $B_{\gamma_1}X(\tau,\sigma)$ and $\tilde{B}_{\gamma_2}X(\tau,\sigma)$
we can find $B_{\gamma_1}\tilde{B}_{\gamma_2}X(\tau,\sigma)$ by reflecting
$X(\tau,\sigma)$ in the middle point of the geodesic interval
connecting $B_{\gamma_1}X(\tau,\sigma)$ and 
$\tilde{B}_{\gamma_2}X(\tau,\sigma)$. Here
$\gamma_1\simeq -1$ and $\gamma_2\simeq 1$.}
% 	\end{center}
\end{figure}

\subsubsection{Properties of B\"acklund transformations.}
We will now describe some properties of the B\"acklund transformations.
It follows from the B\"acklund equations (\ref{LBX}) that
the scalar product of $X(\tau,\sigma)$ and  $B_{\gamma}X(\tau,\sigma)$
is a constant:
\begin{equation}\label{ScalarProduct}
	(X(\tau,\sigma),B_{\gamma}X(\tau,\sigma))=
	{1-\gamma^2\over 1+\gamma^2}
\end{equation}
This follows from considering the scalar product of 
Eqs. (\ref{LBX}) with $B_{\gamma}X$ and $X$.
Taking the scalar product of (\ref{LBX}) with $\partial_+X$ and
$\partial_-X$ and taking into account (\ref{ScalarProduct}),
we can see that the B\"acklund transformations preserve $|\partial_+X|$
and $|\partial_-X|$:
\begin{equation}
	\label{BCommutesWithVirasoro}
	|\partial_{\pm} B_{\gamma}X|=|\partial_{\pm} X|
\end{equation}
An important property (which in our
definition follows from (\ref{LeftContinuous}) and (\ref{RightContinuous}))
is the Permutability Theorem:
\begin{equation}\label{Permutability}
	B_{\gamma_1}B_{\gamma_2}=B_{\gamma_2}B_{\gamma_1}
\end{equation}
This theorem was proven in great generality in 
\cite{Harnad:1983we}\footnote{
In some sense, the permutability theorem is a consequence of the
canonicity (which we will prove in the next subsection) and 
Eq. (\ref{BCommutesWithVirasoro}). Indeed, the ``first Pohlmeyer
charge'' $\int d\sigma |\partial_+X|$, considered as a Hamiltonian,
is ``sufficiently nonresonant'' so that all the Hamiltonians
which commute with it and with the global symmetries
should also commute among themselves. In principle, it should be also possible
to prove the permutability using the arguments based on the
plane wave limit. We notice that the local conserved charges
commute in the plane wave limit, and then use Section
\ref{sec:NoOtherCharges}.}.
An immediate consequence of the commutativity and eqs. (\ref{LBX}) is
that $B_{\gamma_1}B_{\gamma_2}X$ is a linear combination
of $X$, $B_{\gamma_1}X$ and $B_{\gamma_2}X$. The coefficients
of this linear combination can be found from (\ref{ScalarProduct}):
\begin{equation}\label{TangentRule}
	B_{\gamma_1}B_{\gamma_2}X-X=
	{ (X,B_{\gamma_2}X) - (X,B_{\gamma_1}X) \over 
	1-(B_{\gamma_1}X,B_{\gamma_2}X)} (B_{\gamma_1}X-B_{\gamma_2}X)
\end{equation}
This is the ``tangent rule'' of B\"acklund transformations.
For the target space ${\bf R}\times S^2$ the classical string is
essentially equivalent to the sine-Gordon model, and the
tangent rule (\ref{TangentRule})
is closely related to the bilinear identity
for the sine-Gordon $\tau$-function (see \cite{Mikhailov:2005qv}
and references there).

For $\gamma=\tilde{\gamma}$ our definition implies an interesting formula:
\begin{equation}\label{LeftRightRelation}
	B_{\gamma}X=-\tilde{B}_{\gamma}X
\end{equation}
This formula provides relations between the flows of the 
``left'' local charges $\xi_{ {\cal F}_{2n} }$ and the
``right'' local charges $\xi_{ \widetilde{\cal F}_{2n} }$.
We will use these relations in Section \ref{sec:Lift}.
Eq. (\ref{LeftRightRelation}) requires an explanation.
Let us construct the ``perturbative'' solution $B_{\gamma}X$
of (\ref{LBX}) using the null-surface perturbation theory.
Because of the symmetry $X\leftrightarrow B_{\gamma}X$ of 
(\ref{LBX}) there are two such solutions; we will choose
the one which satisfies (\ref{SmallGamma}) when $\gamma$ is small
(the other one should be denoted $B_{-\gamma}X$).
One can see that $-B_{\gamma}X$ satisfies Eq. (\ref{RBX}) for
$\tgamma=\gamma$. Therefore $-B_{\gamma}X$ is either
$\tilde{B}_{\gamma}X$ or $\tilde{B}_{-\gamma}X$, and to verify
the sign we have to verify Eq. (\ref{LargeGamma}).
It is enough to verify Eq. (\ref{LargeGamma}) for 
the null-surface, that is when $\partial_+X=\partial_-X$.
If we consider the null-surface, the solution of the
B\"acklund transformation can be written down explicitly,
see Eqs. (\ref{LeftZerothOrder}) and (\ref{RightZerothOrder}). 
We see that indeed 
$B_{\gamma}X=-\tilde{B}_{\gamma}X$.
But we must stress that (\ref{LeftRightRelation}) is, by our
arguments, valid only in the sector of fast moving strings.
We suspect that (\ref{LeftRightRelation}) might fail outside
of the regime of the fast moving strings because of the
possible problems with the definition of the canonical
$B_{\gamma}$ (such as multivaluedness). In this paper we just
consider the perturbation theory around the null-surface and
use the perturbative definition of $B_{\gamma}$ from Section 
\ref{PerturbativeSolutionsOfBE}.

From (\ref{LeftRightRelation}) and (\ref{TangentRule}) follows the tangent
rule for $B_{\gamma_1}X$ and $\tilde{B}_{\gamma_2}X$:
\[
	B_{\gamma_1}\tilde{B}_{\gamma_2}X+X=
	{ (X,\tilde{B}_{\gamma_2}X) + (X,B_{\gamma_1}X) \over 
	1+(B_{\gamma_1}X,\tilde{B}_{\gamma_2}X)} 
	(B_{\gamma_1}X+\tilde{B}_{\gamma_2}X)
\]
This relation is illustrated on Fig. 1 where $\gamma_1\simeq -1$
and $\gamma_2\simeq 1$.
Equations (\ref{LBX}) and (\ref{RBX})
are symmetric with respect to the exchange $X\leftrightarrow B_{\gamma}X$
and also under $\gamma\to -\gamma$ 
and therefore 
\begin{equation}\label{Inverse}
	B_{-\gamma}=B_{\gamma}^{-1}\;\;\;\mbox{and}\;\;\;
\tilde{B}_{-\tilde{\gamma}}=\tilde{B}_{\tilde{\gamma}}^{-1}
\end{equation}

\subsubsection{Canonicity of B\"acklund transformations.}
\label{sec:Canonicity}
The way we defined B\"acklund transformations 
in Section \ref{sec:Definition}, canonicity follows automatically
from Eqs. (\ref{LeftContinuous}), (\ref{RightContinuous}).
But in fact it can be derived directly from the B\"acklund equations
(\ref{LBX}), (\ref{RBX}). More precisely, if $B_{\gamma}X$ is defined as
the {\em periodic} solution of (\ref{LBX}), then the transformation
$X\mapsto B_{\gamma}X$ is canonical. We will now prove it.

The symplectic form on the classical string phase space can be computed
as the exterior derivative of the ``symplectic potential'' $\alpha$:
\begin{equation}\label{OmegaVsAlpha}
	\omega=\delta\alpha
\end{equation}
Here $\delta$ denotes the differential on the string phase
space. We use $\delta$ instead of the usual notation $d$ for the
differential to distinguish it from the $d$ acting
on the differential forms on the string worldsheet.
The symplectic potential is:
\begin{equation}\label{PresymplecticForm}
	\alpha={1\over 2\pi}\int d\sigma (\delta X, \partial_{\tau}X)
\end{equation}
The symplectic potential is a 1-form on the string phase space,
because it contains one field variation $\delta X$.
Therefore $\omega=\delta\alpha$ is a closed two-form on the
string phase space. The integral in (\ref{PresymplecticForm})
is taken over the spacial contour on the worldsheet at some
fixed $\tau=\tau_0$ and does depend on $\tau_0$. 
However its exterior derivative $\omega$ does not
depend on $\tau_0$. More generally, we could consider a 
family of symplectic potentials depending on the closed
contour $C$ on the worldsheet:
\begin{equation}
	\alpha_C={1\over 2\pi}\oint_C (\delta X, *dX)
\end{equation}
This expression does depend on the choice of the contour $C$,
but its exterior derivative $\omega=\delta \alpha_C$ does not.

Eqs. (\ref{LBX}) and (\ref{ScalarProduct})
imply that the B\"acklund transformation
changes the symplectic potential $\alpha$ by a $\delta$ of something:
\begin{eqnarray}
	B_{\gamma}^*\alpha-\alpha=
	{1\over 2\pi}
	\int d\sigma (\delta B_{\gamma}X, \partial_{\tau}B_{\gamma}X)
	-{1\over 2\pi}
	\int d\sigma (\delta X, \partial_{\tau} X)=\nonumber\\[5pt]
	=\delta\left(\;{1\over 2\pi}
	\int d\sigma (X,\partial_{\sigma}B_{\gamma}X)\right)
	\label{DeltaOfSomething}
\end{eqnarray}
Therefore the symplectic form (\ref{OmegaVsAlpha})
does not change, which
means that  $B_{\gamma}$ is a canonical transformation.

\subsubsection{The ``logarithm'' of the  B\"acklund transformation.}
\label{sec:Logarithm}
Consider the Hamiltonian vector fields $\xi_{\gamma}$ 
and $\tilde{\xi}_{\tgamma}$ defined as the ``logarithms''
of the left and right B\"acklund transformations:
\begin{equation}
	\xi_{\gamma}=\sum_n t_{2n}({\gamma}) \xi_{ {\cal F}_{2n} },\;\;\;
	\tilde{\xi}_{\tgamma}=
	\sum_n \sum_n \tilde{t}_{2n}(\tilde{\gamma})
	\xi_{ \widetilde{\cal F}_{2n} }
\end{equation}
If we define the B\"acklund transformations by Eqs. (\ref{LBX})
and (\ref{RBX}), then $\xi_{\gamma}$ and $\tilde{\xi}_{\tgamma}$
can be defined, at least formally as power series in $\gamma$ and
$\tilde{\gamma}^{-1}$, as follows. 
Let us first define $\xi_{\gamma}^{(1)}$:
\begin{equation}
	\xi_{\gamma}^{(1)}=
	\gamma\left.{\partial\over\partial \gamma}\right|_{\gamma=0}
	B_{\gamma}X
\end{equation}
This is a vector field on the phase space, which depends on $\gamma$ linearly.
Now $e^{-\xi_{\gamma}^{(1)}}B_{\gamma}X=X+O(\gamma^2)$ and therefore
we can define $\xi_{\gamma}^{(2)}$:
\begin{equation}
	\xi_{\gamma}^{(2)}=
	\xi_{\gamma}^{(1)}+{\gamma^2\over 2} 
	\left.{\partial^2\over \partial\gamma^2}\right|_{\gamma=0}
	(e^{-\xi_{\gamma}^{(1)}}B_{\gamma}X)
\end{equation}
We will have $e^{-\xi_{\gamma}^{(2)}}B_{\gamma}X=X+O(\gamma^3)$.
We can continue this process and define
\begin{equation}
	\xi_{\gamma}^{(n+1)}=
	\xi_{\gamma}^{(n)}+{\gamma^{n+1}\over (n+1)!} 
	\left.{\partial^{n+1}\over \partial\gamma^{n+1}}\right|_{\gamma=0}
	(e^{-\xi_{\gamma}^{(n)}}B_{\gamma}X)
\end{equation}
The resulting formal series $\xi_{\gamma}=\xi_{\gamma}^{(\infty)}$
is a vector field on the phase space which could be thought of as
a logarithm of $B_{\gamma}$:
\begin{equation}
	B_{\gamma}=e^{\xi_{\gamma}}
\end{equation}
The vector fields $\xi_{\gamma}$ and $\tilde{\xi}_{\tgamma}$ are power
series in $\gamma$ and $1/\tgamma$. The coefficients of $\gamma$ and 
$1/\tgamma$
are local vector fields on the string phase space. These vector
fields are Hamiltonian, because the B\"acklund transformations are
canonical. Therefore they are generated by some local conserved
charges. We will describe the local conserved charges
in Section \ref{sec:BP} and explain how $\xi_{\gamma}$
and $\tilde{\xi}_{\tgamma}$ are expressed through these
local conserved charges. 
\begin{figure}
	\centering
%	\begin{center}
%\epsfxsize=2.5in {\epsfbox{logarithm.eps}} 
\includegraphics[height=1.6in]{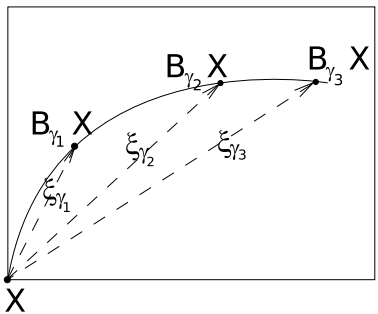}
\caption{$B_{\gamma}$ and $\xi_{\gamma}$.} 
%	\end{center}
\end{figure}

The relation between $B_{\gamma}$ and $\xi_{\gamma}$
is shown schematically on Fig.2.
The solid curve represents the one-parameter family
of solutions $B_{\gamma}X(\tau,\sigma)$ parametrized by
the real $\gamma$. The dashed lines represent the trajectories
of the Hamiltonian vector fields $\xi_{\gamma}$. 
For the finite-gap solutions, all $B_{\gamma}X$ belong to the same
 Liouville torus as $X$. 
The vectors $\xi_{\gamma}$ are the shifts on the torus
corresponding to the B\"acklund transformations. 
But notice that our definition of the ``logarithm'' of the B\"acklund 
transformations did not require the knowledge of the fact that these
transformations
are generated by the flows of the local conserved charges. 
For us here, this fact is rather a consequence of the abstract
definition of the logarithm. Generally speaking, the logarithmic map
from the group of diffeomorphisms to the algebra of vector fields
is notoriously ill-defined. But 
the B\"acklund transformation $B_{\gamma}$ is parametrized by a
continuous parameter $\gamma$, and becomes the identical
transformation at $\gamma=0$. This allows us to define
$\xi_{\gamma}$ as a series in $\gamma$.
In the null surface perturbation theory we have seen that
$B_{\gamma}$ is at each order a rational function of $\gamma$
with the poles at $\gamma=\pm i$. This means that
the series in $\gamma$ converge at least up to $\gamma=1$.

\subsection{Deck transformation and B\"acklund transformations.}
\subsubsection{Deck transformation.}
\label{sec:DeckGeneral}
$AdS_5$ is the universal covering space of the hyperboloid $AdS_5/{\bf Z}$.
In other words, the point $\hat{x}\in AdS_5$ can be thought of as
a point of the hyperboloid $x\in AdS_5/{\bf Z}$, together with the
path connecting $x$ to the fixed ``base point'' $x_*$.
The path is considered modulo the homotopic equivalence (that is,
two pathes are considered equivalent if one can be continuously
deformed to the other). 

The {\em deck transformation} is defined as the isometry of $AdS_5$
which does not change $x$, but changes the path from $x_*$ to $x$
by attaching to it a loop going once around the noncontractible cycle
of the hyperboloid. We will denote the deck transformation $c^2$.

The existence of the large isometry group of $AdS_5$ allows us to express
$c^2$ as a flow by the finite time generated by a Killing vector field.
Namely, if $T$ is the global time of $AdS_5$ then
\begin{equation}
	\label{CViaKillingVector}
	c^2=e^{2\pi {\partial\over \partial T}}
\end{equation}
We defined $c^2$ as an isometry of the AdS space. It can be also considered
a symmetry of the string phase space. Symmetries of the string
theory following from the symmetries of space-time are called
{\em geometrical} symmetries. 

\begin{figure}
	\centering
%	\begin{center}
%	\epsfxsize=2in {\epsfbox{spiral.eps}} 
\includegraphics[height=1.2in]{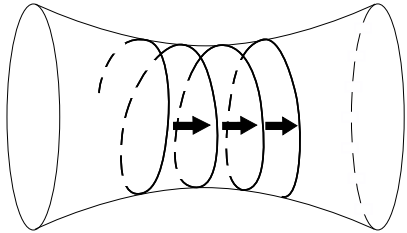}
\caption{The deck transformation measures the deviation of the
string worldsheet from being periodic in the global time.
It is the ``monodromy by the time $2\pi$''. 
The spiral represents the projection of the string worldsheet
to the hyperboloid $AdS_5/{\bf Z}$. To simplify the picture
we have imagined that the string is collapsed to a point.
The noncontractible cycle of the hyperboloid 
corresponds to the time direction in AdS.
} 
%	\end{center}
\end{figure}

Eq. (\ref{CViaKillingVector}) tells us that $c^2$ is what corresponds
in the classical string theory to the notion of the anomalous dimension
in the dual Yang-Mills theory. Additional details can be found in 
\cite{Mikhailov:2004ca}.

The isometry group of $AdS_5$ is $\widetilde{SO}(2,4)$ --- the universal
covering of the orthogonal group. 
Notice that $\partial\over\partial T$ does not commute with the elements
of $\widetilde{SO}(2,4)$ 
(in fact, it is one of the generators of the algebra
$so(2,4)$).  
But $e^{2\pi {\partial\over\partial T}}$ does commute with 
$\widetilde{SO}(2,4)$; it is an element of the center of
$\widetilde{SO}(2,4)$.
Therefore Eq. (\ref{CViaKillingVector}) tells us that for the AdS space,
the deck transformation can be understood as an element of the 
center of the symmetry group $\widetilde{SO}(2,4)$.

We can also interpret Eq. (\ref{CViaKillingVector}) as defining
the ``logarithm'' of the deck transformation:
\[
	"{\partial\over\partial T}={1\over 2\pi}\log c^2 \; "
\]	
But the logarithm is by no means an unambiguously defined operation.
There are other vector fields exponentiating to $c^2$.
In fact, 
when we consider the action of $c^2$ on the classical string phase space,
there is a better choice for $\log c^2$ then 
$2\pi{\partial\over\partial T}$. It is better already because
this $\log c^2$ (unlike ${1\over 2\pi}{\partial\over\partial T}$)
commutes with $\widetilde{SO}(2,4)$; some additional motivation
can be found in \cite{Mikhailov:2004ca}. This alternative definition
of $\log c^2$ relies on the existence of infinitely many local 
conserved charges ${\cal F}_{2k}$.
Remember that we denote $\xi_{ {\cal F}_{2k}}$ the corresponding 
Hamiltonian vector fields.
It turns out that:
\begin{equation}\label{DeckViaF}
	c^2=e^{\sum c_k \xi_{ {\cal F}_{2k} } }
\end{equation}
for some coefficients $c_k$. This does not mean that
${\partial \over\partial T}=\sum c_k \xi_{ {\cal F}_{2k} }$,
but only that the trajectories of the vector field
${\partial \over\partial T}-\sum c_k \xi_{ {\cal F}_{2k} }$
are periodic with the period $1$.
We have conjectured the formula of the type (\ref{DeckViaF})
in \cite{Mikhailov:2004ca}, based on the arguments 
of \cite{Mikhailov:2004au}. We will now give a proof based
on the properties of B\"acklund transformations. 

\subsubsection{Deck transformation is a B\"acklund transformation.}
\label{sec:Lift}
Notice that equations (\ref{LBX}) and (\ref{RBX}) involve only
the projection of the string worldsheet to the hyperboloid $AdS_5/{\bf Z}$.
The string actually lives in $AdS_5$ which is the universal
covering space of this hyperboloid. We can lift the action of the
B\"acklund transformation from the string on the hyperboloid
to the string on $AdS_5$ 
using the vector fields $\xi_{\gamma}$ and 
$\tilde{\xi}_{\tilde{\gamma}}$.

Let us define the continuous families of transformations 
$\bgamma[s]=e^{s\; \xi_{\gamma}}$ and 
$\tbgamma[s]=e^{s\; \tilde{\xi}_{\tgamma}}$
parametrized by a real parameter $s\in [0,1]$.
We do not know if the transformations
$\bgamma[s]$ and $\tbgamma[s]$ have any special properties for 
an arbitrary $s$, 
but for $s=1$ we get the B\"acklund transformations. Now we have a pair of
continuous families of solutions $B_{\gamma}[s]X$ and
$\tilde{B}_{\tgamma}[s]X$ connecting $\bgamma X$
and $\tbgamma X$ to $X$. But the phase space of the classical
string on $AdS$ is a cover of the phase space of the
classical string on the hyperboloid, therefore the existence of
a continuous family of transformations connecting  
the B\"acklund transformation to the identical transformation
allows us to lift it from the string on the hyperboloid to
the string on $AdS_5$.

Now consider the special case when $\gamma=\tilde{\gamma}$.
Eq. (\ref{LeftRightRelation}) implies that for the fast moving strings:
\begin{equation}
	\exp [2(\xi_{\gamma}-\tilde{\xi}_{\gamma})]
	=c^2
\end{equation}
Indeed, Eq. (\ref{LeftRightRelation}) shows that 
the transformation $e^{2(\xi_{\gamma}-\tilde{\xi}_{\gamma})}$
acts on the string on the hyperboloid as the identical transformation.
Therefore it acts on the string in $AdS$ either as the identical transformation
or as some iteration of the deck transformation. We will argue
that this is actually the deck transformation itself.
We know that $e^{2(\xi_{\gamma}-\tilde{\xi}_{\gamma})}=\mbox{(deck)}^k$
where $k$ is some integer. Because of the continuity this integer
cannot change when we continuously change the worldsheet. Therefore 
it is enough to find it for some particular worldsheet. Let us
find $k$ in the limiting case when the fast moving string becomes
a null-surface. In this case the B\"acklund transformations 
are given by (\ref{LeftZerothOrder}) and (\ref{RightZerothOrder}):
\begin{eqnarray}
&&	\bgamma X={1-\gamma^2\over 1+\gamma^2}X
	-{2\gamma\over 1+\gamma^2}{\partial_{\tau}X\over |\partial_{\tau}X|}
	\label{LeftNS}
	\\[5pt]
&&	\tbgamma X={1-\tgamma^{-2}\over 1+\tgamma^{-2}}X
	+{2\tgamma^{-1}\over 1+\tgamma^{-2}}
	{\partial_{\tau}X\over |\partial_{\tau}X|}
	\label{RightNS}
\end{eqnarray}
This means that for the null-surface we have:
\begin{eqnarray}
	&&\xi_{\gamma}=-2{\arctan\gamma\over |\partial_{\tau}X|}
	{\partial\over \partial\tau}
	\label{LeftXiGammaNull}
	\\[5pt]
	&&\tilde{\xi}_{\tgamma}=
	2{\arctan(1/\tgamma)\over |\partial_{\tau}X|}
	{\partial\over\partial\tau}
\end{eqnarray}
For $\gamma=\tilde{\gamma}$:
\begin{equation}
	\xi_{\gamma}-\tilde{\xi}_{\gamma}=
	-\pi{1\over |\partial_{\tau}X|}{\partial\over\partial \tau}
\end{equation}
In the null-surface limit each point of the string moves on the
equator of $AdS_5$, and $e^{s(\xi_{\gamma}-\tilde{\xi}_{\gamma})}$
shifts by the angle $\pi s$ along the equator. Therefore for $s=2$
we go once around the noncontractible cycle of the hyperboloid;
this is the deck transformation.

Let us summarize the relation between the B\"acklund transformations
and the deck transformation. 
There are two continuous families of 
commuting canonical transformations
$B_{\gamma}$ and $\tilde{B}_{\tilde{\gamma}}$. We defined them as power series
in $\gamma$ and $\tilde{\gamma}^{-1}$, so $\gamma$ should be small
and $\tilde{\gamma}$ should be large. But in the fast moving sector
$B_{\gamma}$ and $\tilde{B}_{\tilde{\gamma}}$   can be analytically
continued to finite values of $\gamma$ and $\tilde{\gamma}$. 
To get the deck transformation, we have to consider $\gamma=\tilde{\gamma}$:
\begin{equation}\label{DeckViaBB}
c^2=
B_{\gamma}^2 \tilde{B}_{\gamma}^{-2}
\end{equation}
This suggests that if $Q_{\gamma}$ and $\tilde{Q}_{\tilde{\gamma}}$ are
the quantum operators
corresponding to the B\"acklund transformations in the quantum theory,
then the anomalous dimension on the field theory side 
\cite{Mikhailov:2004ca} is:
\begin{equation}
	e^{2\pi i \Delta E}=Q_{\gamma}^2 \tilde{Q}_{\gamma}^{-2}
\end{equation}
for any parameter $\gamma$.
In this formula it is assumed that $Q_{\gamma}$ is defined first for small 
$\gamma$, and $\tilde{Q}_{\tilde{\gamma}}$ for large $\tilde{\gamma}$,
and then analytically continued to the finite value
$\gamma=\tilde{\gamma}\simeq 1$.

We have argued that $B_{\gamma}$ is an exponential of the
infinite linear combination of the local charges. 
Therefore (\ref{DeckViaBB}) implies (\ref{DeckViaF}).

\section{Local Pohlmeyer charges.}
\label{sec:BP}
\subsection{The definition of Pohlmeyer charges.}
\label{sec:DefinitionOfCharges}
The Pohlmeyer charges are the local conserved charges
of the sigma-model. 
There are  two series of these charges,
which we call ``left charges'' ${\cal F}_{2k}$
and ``right charges'' $\tilde{\cal F}_{2k}$.
They can be obtained from the B\"acklund
transformation. 
The generating function of the Pohlmeyer charges is:
\begin{equation}\label{LeftGenerating}
	{\cal F}_{left}(\gamma)={1\over 2\pi}\int d\sigma 
\left[\gamma(\bgamma X,\partial_+X)+\gamma^3(\bgamma X,\partial_-X)\right]
\end{equation}
\begin{equation}\label{RightGenerating}
	{\cal F}_{right}(\tgamma)={1\over 2\pi}\int d\sigma 
	\left[{1\over \tgamma}
	(\tbgamma X,\partial_-X)+{1\over\tgamma^3}(\tbgamma X,\partial_+X)\right]
\end{equation}
Here $B_{\gamma}X$ and $\tilde{B}_{\tilde{\gamma}}X$ are defined
as perturbative solutions of Eqs.
(\ref{LBX}), (\ref{RBX}), (\ref{SmallGamma}), (\ref{LargeGamma}).
The analogous formulas can be written for the Pohlmeyer charges
of the $S^5$ part of the sigma-model.

The generating functions ${\cal F}_{left}(\gamma)$ and 
${\cal F}_{right}(\tgamma)$ have an expansion 
in the even powers\footnote{The symmetry 
${\cal F}_{left}(\gamma)={\cal F}_{left}(-\gamma)$
follows from (\ref{Inverse}) and (\ref{ScalarProduct})
and the fact that the B\"acklund transformations preserve ${\cal F}_{left}$.} 
of $\gamma$ and $\tgamma^{-1}$ respectively. The coefficients are
the local conserved charges ${\cal F}_{2k}, \tilde{\cal F}_{2k}$. 
But we can also solve the B\"acklund equations in the
null-surface perturbation theory, using $1\over |\partial_{\tau}X|$
as a small parameter. When we solve the B\"acklund equations in the
null-surface perturbation theory, each order of the expansion
of $B_{\gamma}X$ in $1\over |\partial_{\tau}X|$ depends on $\gamma$
as a rational function. This means that 
at each order of the null-surface perturbation theory 
${\cal F}_{left}(\gamma)$ and ${\cal F}_{right}(\tilde{\gamma})$ are rational
functions of $\gamma$ and $\tilde{\gamma}$ respectively.
Therefore in the null-surface perturbation theory 
${\cal F}_{left}(\gamma)$ and ${\cal F}_{right}(\tilde{\gamma})$ 
have a well-defined analytic continuation to finite values of
$\gamma$ and $\tilde{\gamma}$.

It was conjectured in \cite{Arutyunov:2003rg}\footnote{The improved
currents of \cite{Arutyunov:2003rg} are the sums of left
and right charges; we use a slightly different version
of improved charges here, which involve only the left ${\cal F}_{2k}$.} that
there exist the ``improved'' charges ${\cal G}_{2k}^{left}, 
{\cal G}_{2k}^{right}$
which are finite linear combinations 
of ${\cal F}_{2k}$ and $\tilde{\cal F}_{2k}$:
\begin{equation}
	{\cal G}_{2k}^{left}=\sum_{n=1}^k c_{k,n} {\cal F}_{2k}
\end{equation}
with the property that in the null-surface perturbation theory
\begin{equation}\label{DefImproved}
	{\cal G}_{2k}^{left}\simeq |\partial_{\tau} X|^{-k+1}
	\;\;\; (k\geq 2)
\end{equation}
The null-surface expansion of the generating functions (\ref{LeftGenerating})
and (\ref{RightGenerating})
is the expansion in the ``improved'' charges ${\cal G}^{left}_{2k}$.
The coefficients of ${\cal G}^{left}_{2k}$ in 
the expansion of ${\cal F}_{left}(\gamma)$ are rational 
functions of $\gamma$. (Indeed, if we solve the B\"acklund
equations in the null-surface perturbation theory, each
order will depend on $\gamma$ as a rational function.)
We will derive the explicit formula (\ref{GeneratingImproved})
using the plane wave limit (but our derivation of
(\ref{GeneratingImproved}) works only under
the assumption that the improved charges satisfying (\ref{DefImproved})
exist).

We will now explain how the vector field $\xi_{\gamma}$ can be
represented as the linear combination of the Hamiltonian vector
fields generated by ${\cal F}_{2k}$ and $\widetilde{\cal F}_{2k}$.
We already know that $\xi_{\gamma}$ is represented by a linear
combination of the Hamiltonian vector fields generated
by the local conserved charges, see Section 
\ref{sec:Logarithm}. We will argue
in Section \ref{sec:NoOtherCharges} that there are no local conserved
charges other than Pohlmeyer's. 
Therefore $\xi_{\gamma}$ is a linear combination of the Pohlmeyer
charges. We will fix the coefficients of this linear combination
in Section \ref{sec:Generator}.

\subsection{The null-surface limit of $\xi_{\gamma}$.}
\label{sec:NullSurfaceLimit}
In the null surface limit we have $|\partial_{\tau}X|>> |\partial_{\sigma}X|$
and therefore we can replace in Eq. (\ref{LeftGenerating}) 
$\partial_{\pm} X\simeq {1\over 2} \partial_{\tau}X$.
We get:
\begin{equation}
	{\cal F}_{left}(\gamma)\simeq {1\over 4\pi}\int d\sigma 
	\gamma(1+\gamma^2)(\bgamma X,\partial_{\tau}X)
	\simeq -{\gamma^2\over 2\pi}\int d\sigma |\partial_{\tau}X|
\end{equation}
where we have also used (\ref{LeftNS}).
The symplectic structure of the classical string is:
\begin{equation}
	\omega={1\over 2\pi}\int d\sigma (\delta \partial_{\tau} X,\delta X)
\end{equation}
Therefore in the null-surface limit the functional ${\cal F}_{left}(\gamma)$
generates the following Hamiltonian vector field\footnote{Given a function
$H$ on the string phase space, we denote $\xi_H$ the corresponding
Hamiltonian vector field, $\iota_{\xi_H}\omega=-\delta H$. 
This is the usual notation in the classical mechanics.
On the other hand, we denote $\xi_{\gamma}$ the vector field which is
the ``logarithm'' of $B_{\gamma}$. We hope that this
will not cause a confusion.}:
\begin{equation}\label{XiFZerothOrder}
	\xi_{{\cal F}_{left}(\gamma)}.X=
	-\gamma^2{1\over |\partial_{\tau}X|}\partial_{\tau}X
\end{equation}
On the other hand (\ref{LeftXiGammaNull}) gives us the formula for
$\xi_{\gamma}$ in the null-surface limit:
\begin{equation}\label{XiGammaZerothOrder}
\xi_{\gamma}=-2{\arctan\gamma\over |\partial_{\tau}X|}
	{\partial\over \partial\tau}
\end{equation}
The null-surface limit is too strict and does not
provide us enough information about $\xi_{\gamma}$
to determine its expansion in $\xi_{ {\cal F}_{2k}}$. 
We will see that the plane wave limit is good enough for this purpose.
In Section \ref{sec:Generator} we will derive the formula
relating $\xi_{\gamma}$ to $\xi_{ {\cal F}(\gamma) }$:
\begin{equation}\label{FromNSL}
	\xi_{\gamma}=2\int_0^{\gamma} {d\gamma\over \gamma^2(1+\gamma^2)}
	\xi_{{\cal F}_{left}(\gamma)}
\end{equation}
As we have already discussed, $\xi_{ {\cal F}_{left}(\gamma)}$ is at each order
of the null-surface perturbation theory a rational function of $\gamma$.
For example, the zeroth approximation given by (\ref{XiFZerothOrder})
depends on $\gamma$ as $\gamma^2$. 
Because of the integral transform (\ref{FromNSL}),
$\xi_{\gamma}$ is not a rational function
of $\gamma$. For example the zeroth order (\ref{XiGammaZerothOrder})
contains $\arctan \gamma$. But after the exponentiation 
 $B_{\gamma}=e^{\xi_{\gamma}}$ becomes again a rational
function in the null-surface expansion, for example the zeroth order
is given by (\ref{LeftNS}):
\[
\bgamma X={1-\gamma^2\over 1+\gamma^2}X
	-{2\gamma\over 1+\gamma^2}{\partial_{\tau}X\over |\partial_{\tau}X|}
\]

\subsection{The plane wave limit.}
In the plane wave limit \cite{BMN} the classical string
becomes essentially the theory of free massive field.
The Pohlmeyer charges in this limit can be written explicitly. 
It turns out that all the local conserved charges
of the free massive field invariant under the global symmetry 
appear as the limits of the Pohlmeyer charges.
Also, the Pohlmeyer charges remain linearly independent in the
plane wave limit.
 This implies
that all the local charges appearing in the expansion of $\xi_{\gamma}$
are Pohlmeyer charges, and allows to find the coefficients of the
expansion. This gives Eq. (\ref{Spectrality}).
We then give a direct derivation of (\ref{Spectrality}) without
using the plane wave limit.

\subsubsection{Pohlmeyer charges in the plane wave limit.}
\label{sec:PohlmeyerChargesInPWL}
In the plane wave limit the motion of the string is 
restricted to the small neighborhood of the null-geodesic.
The transverse components of $X$ become free fields
$x^I$, $I=1,2,3,4$. We have $X^{1,2,3,4}=\epsilon x^{1,2,3,4}+o(\epsilon)$,
where $\epsilon$ is a small parameter measuring the accuracy of the
plane wave approximation.
The longitudinal components are:
\begin{eqnarray}
&&	x_+=M(\tau-\tau_0)-{\epsilon^2\over 2}x_-\\[5pt]
&&	\partial_{\tau}x_-=-{1\over 2 M}\sum_{i=1}^4
	\left[(\partial_{\tau}x_i)^2+(\partial_{\sigma}x_i)^2-M^2x_i^2
	\right]\\[5pt]
&&	\partial_{\sigma}x_-=-{1\over M}\sum_{i=1}^4
	(\partial_{\tau}x_i,\partial_{\sigma}x_i)
\end{eqnarray}
Here $M=\epsilon^2J+\ldots$ where $J$ is the angular momentum, 
see the review section of \cite{Mikhailov:2005wn} for details.
The free equations of motion for $x^I$ are
$(\partial_{\tau}^2-\partial_{\sigma}^2+M^2)x^I=0$.

Let $\alpha^I_n$ be 
the Fourier modes\footnote{The index $n$ does not have to be an integer,
because the periodicity of the $\sigma$-coordinate is not important
for our arguments. In the following formulas we use
the summation over $n$ assuming the periodic boundary
conditions; for an infinite worldsheet we would have
to replace the summation over $n$ with the integration.}
of $x^I$.
The generating function of the local conserved charges
in the plane wave limit was computed in 
\cite{Mikhailov:2005wn}\footnote{The difference with \cite{Mikhailov:2005wn} 
in the overall factor
of $\epsilon^2$ is due to a different normalization;
also notice that in \cite{Mikhailov:2005wn} we discussed the 
$S^5$ part of the string worldsheet, and because of that
the formulas contained $J$ instead of $E$.}:
\begin{eqnarray}
	{\cal F}_{left}(\gamma)=
	-\gamma^2\left[E+(1+\gamma^2)
	\sum\limits_{n=-\infty}^{\infty} \sum\limits_{I=1}^4
	{\pp(\omega_n+ n)\over \pp^2+\gamma^2(\omega_n+ n)^2}
	\alpha^I_n\overline{\alpha^I_n}\right] =\nonumber\\[5pt]=
	-\tan^2\alpha\left[E+
	\sum\limits_{n=-\infty}^{\infty} \sum\limits_{I=1}^4
	{\pp^{-1}(\omega_n + n)\over 1+
	\sin^2\alpha[\pp^{-2}(\omega_n+ n)^2-1]}
	\alpha^I_n\overline{\alpha^I_n}\right]
\end{eqnarray}
Here $E$ is the energy (corresponding to $\partial\over\partial T$),
$\omega_n=\sqrt{M^2+n^2}$ and  $\alpha$ is defined as follows:
\begin{equation}
	\gamma=\tan\alpha
\end{equation}

\subsubsection{Improved charges.}
The oscillator expansion of ${\cal G}^{left}_{2k}$ is 
\begin{equation}
	{\cal G}^{left}_{2k} =
	\delta_{k,1} E+
	\sum\limits_{n=-\infty}^{\infty}
	\sum\limits_{I=1}^4 
	{\omega_n + n\over \pp}
	\left[{(\omega_n+ n)^2\over \pp^{2}}-1\right]^{k-1}
	\alpha^I_n\overline{\alpha^I_n}
\end{equation}
This means that the coefficients of the expansion of 
${\cal F}^{left}(\gamma)$ are rational functions of $\gamma$:
\begin{equation}\label{GeneratingImproved}
	{\cal F}_{left}(\gamma)=
	\tan^2\alpha
	\sum\limits_{k=1}^{\infty}(-1)^{k}(\sin\alpha )^{2k}\;
	{\cal G}^{left}_{2k}=
	\sum\limits_{k=1}^{\infty}(-1)^{k}{\gamma^{2k+2}\over (1+\gamma^2)^k}\;
	{\cal G}^{left}_{2k}
\end{equation}
We derived this formula in the plane wave limit, but 
the following statement should be 
true for general fast moving strings:
\begin{quote}
	The local charges ${\cal G}^{left}_{2k}$
		related to the generating function
		${\cal F}_{left}(\gamma)$ by Eq. (\ref{GeneratingImproved})
		are improved in a sense of Eq. (\ref{DefImproved}).
\end{quote}
Indeed, if we assume the existence of the ``improved'' charges
characterized by (\ref{DefImproved}) then the coefficients
of these charges in the expansion of ${\cal F}_{left}(\gamma)$
are fixed unambiguously by the plane wave limit.

\subsubsection{There are no local conserved charges other than Pohlmeyer's
and the charges corresponding to the global symmetries.}
\label{sec:NoOtherCharges}
Given a local conserved charge, we can consider its leading
term  in the plane wave limit.
It will be of the form
\begin{equation}\label{LeadingOrderChargePWL}
	\int d\sigma \;
	\epsilon^q M^{\left(1-\sum_{j=1}^q k_j\right)}
	\partial^{k_1} x^{i_1}\cdots
	\partial^{k_q} x^{i_q}
\end{equation}
The power of $M$ is found from the worldsheet scaling invariance.
But the local conserved currents of the free field theory are
all quadratic in oscillators, and therefore the
local charge should have $q=0$ or $q=2$.
These are either Pohlmeyer charges or global conserved charges.

If a nonzero conserved charge
has a leading expression in the plane wave limit of the order $\epsilon^3$
or higher, then it is necessarily a nonlocal charge. This is because
all the local conserved charges of the free massive field are quadratic
in the oscillators, and therefore such a conserved charge would be
nonlocal already in the plane wave limit.

Therefore a local conserved charge is completely fixed by its plane
wave limit, and also all the local conserved charges are Pohlmeyer 
charges.

\subsubsection{Pohlmeyer charges and quasimomenta.}
The Pohlmeyer charges can be also obtained from the spectral 
invariants  of the monodromy matrix.
The precise relation between the Pohlmeyer charges and the
quasimomenta for the classical string
on $S^5$ was derived in \cite{Arutyunov:2005nk} using the
approach of \cite{Harnad:1983we}:
\begin{equation}\label{AZ}
	\begin{array}{rl}
	{\cal F}_{left}(\gamma)=-{i\over 2\pi}{\gamma^3 \over 1+\gamma^2}
	&\left[p_1\left({1-i\gamma\over 1+i\gamma}\right)+
	p_2\left({1-i\gamma\over 1+i\gamma}\right)+\right.\\[5pt]&\left.
	+p_3\left({1+i\gamma\over 1-i\gamma}\right)+
	p_4\left({1+i\gamma\over 1-i\gamma}\right)\right]
	=\\[5pt]={1\over\pi}{\gamma^3 \over 1+\gamma^2}
	&\mbox{Im}\left[p_1\left({1-i\gamma\over 1+i\gamma}\right)+
	p_2\left({1-i\gamma\over 1+i\gamma}\right)\right]
\end{array}
\end{equation}
The quasimomenta $p_1,p_2,p_3,p_4$ are the logarithms of 
the eigenvalues of the monodromy matrix 
\cite{Kazakov:2004qf,Kazakov:2004nh,Beisert:2004ag,Beisert:2005bm,%
Schafer-Nameki:2004ik}.

In Section 4.2  of \cite{Mikhailov:2005wn}
we conjectured the integral formula for the energy shift:
\begin{equation}\label{IntegralFormulaForHInfty}
	\Delta E=-{1\over \pi}\int_{-\infty}^{\infty}
	{d\gamma \over \gamma^2 (1+\gamma^2)}
	{\cal F}_{left}(\gamma)
\end{equation}
Eq. (\ref{GeneratingImproved}) shows that this formula
encodes the expansion of $\Delta E$ as an infinite linear combination
of the improved currents ${\cal G}^{left}_{2k}$.
Now let us substitute the result (\ref{AZ}) of \cite{Arutyunov:2005nk} 
and change the integration variable $\gamma\to y={1-i\gamma\over 1+i\gamma}$.
We get:
\begin{equation}
	\Delta E =	-{i\over 16\pi^2}\int\limits_{|y|=1}
	d\left(y+{1\over y}\right)
	[p_1(y)+p_2(y)+p_3(\overline{y})+p_4(\overline{y})]
\end{equation}
Since $|y|=1$ we can rewrite this integral in the following way:
\begin{equation}
	\Delta E =	-{i\over 16\pi^2}
	\int\limits_{|y|=1} d\left(y+{1\over y}\right)
	\sum\limits_{j=1}^4 \varepsilon_j p_j(y)
\end{equation}
where $\varepsilon_{1,2}=1$ and $\varepsilon_{3,4}=-1$.
The contour of integration is the circle $|y|=1$.
Similar formulas for the energy shift were obtained 
in \cite{Kazakov:2004nh,Beisert:2004ag,Beisert:2005bm}.
The contour can be deformed, in particular it does not have
to go through the points $y=\pm 1$. But if we want to deform the
contour, we should make sure that it does not cross 
the singularities of ${\cal F}_{left}$ which for the fast moving strings
are located near $y=0,\infty$ (see Section \ref{PerturbativeSolutionsOfBE};
in the null-surface expansion all the poles are at $\gamma=\pm i$).

We will now give a proof of Eq. (\ref{IntegralFormulaForHInfty}).

\subsubsection{Generator of B\"acklund transformations.}
\label{sec:Generator}
The action of B\"acklund transformations on the oscillators
in the plane wave limit
was derived in \cite{Mikhailov:2005wn}:
\begin{equation}
	\bgamma \alpha_n={\pp-i\gamma(\omega_n+n)\over 
	\pp+i\gamma(\omega_n+n)}\alpha_n
\end{equation}
where $\pp$ is the mass parameter of the plane wave theory
(which we denoted $\mu$ in \cite{Mikhailov:2005wn}) 
and $\omega_n=\sqrt{\pp^2+n^2}$.
The coordinates $x_+$ (which corresponds to the time in the
light-cone gauge) and $x_-$ transform under the B\"acklund transformations
as follows:
$$
	x_+\to x_+-\arctan\gamma,\;\;\;
	x_-\to x_-+{2\over\epsilon^2}\arctan\gamma
$$
This means that
\begin{eqnarray}
&&	\xi_{\gamma}\alpha_n=\log \left[ {\pp-i\gamma(\omega_n+n)\over 
	\pp+i\gamma(\omega_n+n)}\right]\alpha_n 
	\\[5pt]
&&	\xi_{\gamma}x_+=-\arctan\gamma
	\\[5pt]
&&	\xi_{\gamma}x_-={2\over \epsilon^2}\arctan\gamma
\end{eqnarray}	
Therefore the Hamiltonian $H_{\gamma}$ of $\xi_{\gamma}$ is:
\begin{equation}\label{HGammaInPWL}
	H_{\gamma}=-2\arctan(\gamma)E-i\sum_n
	\log \left[ {\pp-i\gamma(\omega_n+n)\over 
	\pp+i\gamma(\omega_n+n)}\right]\alpha_n \overline{\alpha_n}
\end{equation}
where $E$ generates 
$-{1\over \epsilon^2}{\partial\over\partial x_-}+
{1\over 2}{\partial\over\partial x_+}$.
On the other hand the generating function of the Pohlmeyer
charges was also computed in \cite{Mikhailov:2005wn}:
\begin{equation}\label{FLeftGammaInPWL}
	{\cal F}_{left}(\gamma)=
	-\gamma^2\left[E+(1+\gamma^2)\sum_n
	{\pp(\omega_n+ n)\over \pp^2+\gamma^2(\omega_n+ n)^2}
	\alpha_n\overline{\alpha_n}\right]
\end{equation}
We see that the Pohlmeyer charges ${\cal F}_{2n}$
(which are the coefficients
of $\gamma^{2n}$ in the expansion of ${\cal F}_{left}(\gamma)$)
are linearly independent in the plane  wave limit. 
Therefore the comparison of (\ref{HGammaInPWL}) and
(\ref{FLeftGammaInPWL}) allows us to fix the coefficients
in the expansion of $H_{\gamma}$ in ${\cal F}_{2n}$.
The result can be written in the following compact form:
\begin{equation}\label{Spectrality}
	{\partial H_{\gamma}\over \partial \gamma}=
	{2\over \gamma^2(1+\gamma^2)}{\cal F}_{left}(\gamma)
\end{equation}
We call $H_{\gamma}$ the ``generator'' of B\"acklund transformations.
Eq. (\ref{Spectrality}) relates $H_{\gamma}$ to the generating function
of Pohlmeyer charges. 
It is closely related to the  {\em spectrality} of the B\"acklund
transformations which is explained in \cite{SklyaninBQ}.
We will give a direct derivation of this equation in the next subsection.

Eq.  (\ref{Spectrality}) implies that
 $H_{\gamma}$ can be represented 
as an  integral over the open contour:
\begin{eqnarray}
&&	H_{\gamma}=2\int_0^{\gamma} {d\gamma\over \gamma^2(1+\gamma^2)}
	{\cal F}_{left}(\gamma)=
	\label{ExplicitFormula}\\[5pt]
	&&	={i\over 8\pi}\int^{y={1-i\gamma\over 1+i\gamma}}
	dy\left({1\over y^2}-1\right)
	[p_1(y)+p_2(y)+p_3(\overline{y})+p_4(\overline{y})]
	\nonumber
\end{eqnarray}
Here the integration contour in the $y$-plane starts at $y=1$
and goes along the circle $|y|^2=1$ to the point 
$y={1-i\gamma\over 1+i\gamma}$.
Since $|y|=1$ we have:
\begin{eqnarray}
&&	H_{\gamma}=
-{i\over 16\pi}
\int\limits_{y={1+i\gamma\over 1-i\gamma}}^{y={1-i\gamma\over 1+i\gamma}}
	d\left(y+{1\over y}\right)
	\sum\limits_{j=1}^4\varepsilon_j p_j(y)
	\label{OpenContour}
\end{eqnarray}
\begin{figure}
	\centering
%	\begin{center}
%	\epsfxsize=2.1in {\epsfbox{contours.eps}} 
\includegraphics[height=2in]{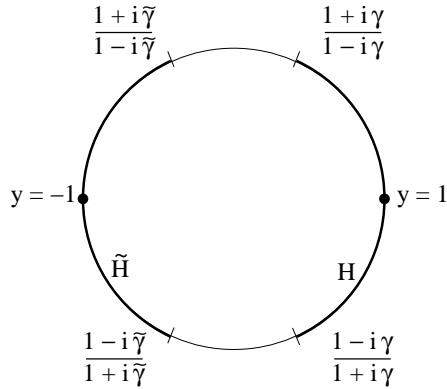}
\caption{The contours for $H_{\gamma}$ and $\tilde{H}_{\tilde{\gamma}}$.}
%	\end{center}
\end{figure}
\noindent
The contour goes along the circle $|y|=1$ through the point $y=1$.
We see that the B\"acklund transformations are generated
by the same integral as the deck transformation, 
but the contour is open.
The positions of the endpoints depend on the parameter
$\gamma$ of the B\"acklund transformation.
When $\gamma$ is small we can compute this integral by expanding
$\sum \varepsilon_j p_j(y)$ around $y=1$; this again tells us
that the B\"acklund transformations are generated by the
local conserved charges. 

In Section (\ref{PerturbativeSolutionsOfBE}) we have seen that
$B_{\gamma}X$ is a rational function of $\gamma$ in the null-surface
perturbation theory, with the poles at $\gamma=\pm i$.
The same is true about ${\cal F}_{left}(\gamma)$, because
${\cal F}_{left}(\gamma)$ is related to $B_{\gamma}X$ 
by Eq. (\ref{LeftGenerating}).
Therefore, if we consider ${\cal F}_{left}(\gamma)$ as an
expansion in powers of $\gamma$ near $\gamma=0$, then 
at each order of the null-surface expansion the series
in $\gamma$ will converge at least up to $\gamma=1$.
Eq. (\ref{ExplicitFormula}) tells us that
$H_{\gamma}$ as a power series in $\gamma$ 
also converges in the null-surface expansion.
Therefore, although our arguments for $B_{\gamma}=e^{\xi_{\gamma}}$
were based on the expansion in powers of $\gamma$, in fact
this relation between $B_{\gamma}$ and $\xi_{\gamma}$
should be true in the
null-surface expansion in some finite region of the parameter $\gamma$.

The generator of the right B\"acklund transformation 
$\tilde{H}_{\tilde{\gamma}}$ is given by the same formula 
as (\ref{OpenContour}) except for the contour goes through
$y=-1$.
When $\gamma=\tgamma$ the difference $H_{\gamma}-\tilde{H}_{\gamma}$
is given by the integral over the closed
contour.
Notice that for the fast moving string
$H_{\gamma}-\tilde{H}_{\gamma}$
does not depend on $\gamma$.
This is because the conserved charge
generating the deck transformation 
(or the action variable \cite{Mikhailov:2004au}) 
cannot be deformed.  
It is natural to denote $H_{\gamma}-\tilde{H}_{\gamma}=H_{\infty}$.
Eq. (\ref{DeckViaBB}) implies that $H_{\infty}=\pi \Delta E$.
Therefore we derived (\ref{IntegralFormulaForHInfty}).

The integral formula (\ref{OpenContour}) involves a particular
combination of the quasimomenta $p_1,p_2,p_3,p_4$. 
If we integrate other combinations of the quasimomenta,
we will get some other Hamiltonians, which will
be non-local even in the null-surface limit.
In the plane wave limit these other combinations should
give the conservation laws of the qubic and higher
order in the oscillators (the local Pohlmeyer charges
are quadratic).
It would be interesting to see if they
have any geometrical meaning like B\"acklund transformations.

\subsection{A direct derivation of (\ref{Spectrality}).}
Here we will derive Eq. (\ref{Spectrality})
without the use of the plain wave limit. 
The  main ingredients are the canonicity of B\"acklund
transformations (\ref{DeltaOfSomething}) and the permutability 
(\ref{Permutability}) which implies
that the conserved charges are invariant under the B\"acklund
transformations. 
We will use an interpretation of a canonical transformation
$(p,q)\mapsto (\tilde{p},\tilde{q})$ as defining
a Lagrangian submanifold in the direct product
of two copies of the phase space with the symplectic form
$dp\wedge dq -d\tilde{p}\wedge d\tilde{q}$.
(This Lagrangian submanifold is the graph of the canonical
transformation.)

Let us fix a contour on the string worldsheet at $\tau=0$.
The phase space is parametrized by the contours $X(\sigma)$
and $\partial_{\tau}X(\sigma)$ at $\tau=0$.
Since $B_{\gamma}$ is a canonical transformation,
the graph of $B_{\gamma}$ is a Lagrangian submanifold in
the direct product of two phase spaces.
We will call it $L_{\gamma}$.
By definition $L_{\gamma}$ consists of pairs $(x,B_{\gamma}x)$,
where $x$ is a point of the string phase space ($x$ is a pair 
$X(\sigma), \partial_{\tau}X(\sigma)$).
Let us consider two points in the phase space,
$x^{(1)}$ and $x^{(2)}$, and their B\"acklund transforms
$B_{\gamma}x^{(1)}$ and $B_{\gamma}x^{(2)}$.

Let us integrate the 1-form $B_{\gamma}^*\alpha-\alpha$ introduced in
Section \ref{sec:Canonicity} over a contour in $L_{\gamma}$ connecting
$(x^{(1)},B_{\gamma}x^{(1)})$ to $(x^{(2)},B_{\gamma}x^{(2)})$.
The integral will not depend on the choice of the contour,
because $L_{\gamma}$ is a Lagrangian manifold.
Let us denote the integral $I$. Eq. (\ref{DeltaOfSomething}) implies:
\begin{equation}\label{FormulaForI}
I=\left.-{1\over 2\pi}\int d\sigma (\partial_{\sigma}X,B_{\gamma}X)
\right|_1^2
\end{equation}
where $F|_1^2$ means 
$F(x^{(2)})-F(x^{(1)})$.
Let us now consider
an infinitesimal variation of $\gamma$. On the new Lagrangian
manifold $L_{\gamma+\delta\gamma}$ consider two points
$(x^{(1)},B_{\gamma+\delta\gamma}x^{(1)})$ and
$(x^{(2)},B_{\gamma+\delta\gamma}x^{(2)})$, connect them by a contour
lying in $L_{\gamma+\delta\gamma}$ and consider the new value of the 
integral $I$.
Let us calculate the variation of $I$:
\begin{eqnarray}
&&	{\partial I\over\partial\gamma} = \left.{1\over 2\pi}\int d\sigma 
	\left(\partial_{\tau}B_{\gamma}X,
	\left.{\partial B_{\gamma}X
	\over\partial\gamma}\right|_{X,\partial_{\tau}X}
	\right)\right|_1^2+
	\int_1^2\iota_{\partial_{\gamma}\xi_{\gamma}}\omega=
	\nonumber
	\\[5pt]
&&	=\left.-{1\over 2\pi}\int d\sigma 
	\left(\partial_{\sigma}X,
	\left.{\partial B_{\gamma}X\over\partial\gamma}
	\right|_{X,\partial_{\tau}X}\right)\right|_1^2
	\label{VariationOfI}
\end{eqnarray}
We have used Eq. (\ref{FormulaForI}). The vertical
bar indicates that we differentiate with respect to $\gamma$
with the constant
$X$ and $\partial_{\tau}X$ (that is, only $B_{\gamma}X$ and
$\partial_{\tau}B_{\gamma}X$ changes). 
\begin{figure}
	\centering
%	\begin{center}
%	\epsfxsize=2.5in {\epsfbox{graph.eps}} 
\includegraphics[height=2.2in]{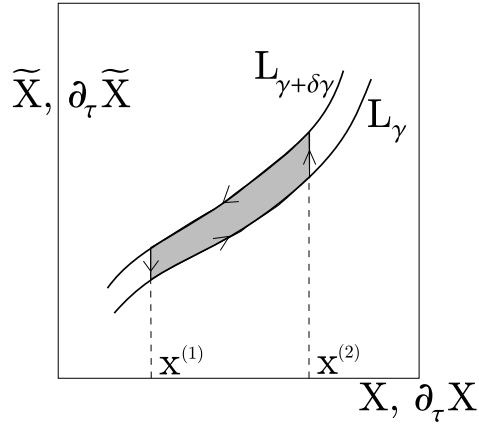}
\caption{Because of the Stokes' theorem the contour integral of the 1-form
$\int d\sigma ((\delta X,\partial_{\tau}X)-
(\delta \widetilde{X},\partial_{\tau}\widetilde{X}))$ is equal to the
integral over the strip of the 2-form 
$\int d\sigma ((\delta X,\partial_{\tau}\delta X)-
(\delta \widetilde{X},\partial_{\tau}\delta \widetilde{X}))$}
\label{StokesTheorem}
%\end{center}
\end{figure}
The term
$\int_1^2\iota_{\partial_{\gamma}\xi_{\gamma}}\omega$
on the left hand side appears when we integrate $\omega$
over the narrow strip between the path on $L_{\gamma}$ going from
$(x^{(1)},B_{\gamma}x^{(1)})$ to $(x^{(2)},B_{\gamma}x^{(2)})$
and the path  on $L_{\gamma+\delta\gamma}$ going from 
$(x^{(1)},B_{\gamma+\delta\gamma}x^{(1)})$ to 
$(x^{(2)},B_{\gamma+\delta\gamma}x^{(2)})$, see Fig. \ref{StokesTheorem}.
It is equal to:
\begin{equation}\label{IntegralOverStrip}
	\int_1^2\iota_{\partial_{\gamma}\xi_{\gamma}}\omega=
	-{\partial H_{\gamma}\over\partial\gamma}(x^{(2)})+
	{\partial H_{\gamma}\over\partial\gamma}(x^{(1)})
\end{equation}
Here
we have used the permutability $B_{\gamma_1}B_{\gamma_2}=
B_{\gamma_2}B_{\gamma_1}$ which implies that 
\begin{equation}\label{InvarianceOfH}
H_{\gamma_1}(B_{\gamma_2}x)=H_{\gamma_1}(x)
\end{equation}
In other words,
the local conserved charges are preserved by the B\"acklund transformation.
We have: $\partial_{\gamma}B_{\gamma}x=\partial_{\gamma}
(e^{\xi_{\gamma}}.x)$. Because of the permutability this is equal to 
$(\partial_{\gamma_1}\xi_{\gamma_1}|_{\gamma_1=\gamma}).B_{\gamma}x$,
in other words the value of the vector field $\partial_{\gamma}\xi_{\gamma}$
at the point $B_{\gamma}x$.
And $\iota_{\partial_{\gamma}\xi_{\gamma}}\omega$ at the
point $B_{\gamma}x$ is equal to $-d\partial_{\gamma}H_{\gamma}$ 
at the point $B_{\gamma}x$
(this is an element of the cotangent space to the phase space
at the point $B_{\gamma}x$). Therefore the integral over the strip is
equal to  
\[
-\left.{\partial\over\partial\gamma_1}\right|_{\gamma_1=\gamma}
H_{\gamma_1}(B_{\gamma}x_2)+
\left.{\partial\over\partial\gamma_1}\right|_{\gamma_1=\gamma}
H_{\gamma_1}(B_{\gamma}x_1)
\]
and because of (\ref{InvarianceOfH}) this is equal to the right hand
side of (\ref{IntegralOverStrip}).
Combining Eqs. (\ref{IntegralOverStrip}) and  (\ref{VariationOfI}) we get:
\[
	\left.{\partial H_{\gamma}\over\partial \gamma}
	\right|_{X,\; \partial_{\tau}X}
	={1\over 2\pi}\int d\sigma \left[ \left( \partial_{\sigma}X,
	\left.{\partial B_{\gamma}X 
	\over\partial \gamma}\right|_{X,\; \partial_{\tau}X}
	\right)
	+\left(\partial_{\tau}B_{\gamma}X, 
	\left.{\partial B_{\gamma}X\over\partial \gamma}
	\right|_{X,\;\partial_{\tau}X}\right)
	\right]
\]
Now let us substitute $\partial_{\sigma}X+\partial_{\tau}B_{\gamma}X$
from the B\"acklund equation (\ref{LBX}):
\begin{equation}
	\partial_{\tau}B_{\gamma}X+\partial_{\sigma}X=
	-{1\over 2}(1+\gamma^2)(B_{\gamma}X,\partial_-X)X
	-{1\over 2}(1+\gamma^{-2})(B_{\gamma}X,\partial_+X)X+\ldots
\end{equation}
where dots denote terms proportional to $B_{\gamma}X$. 
Notice that $(X,\partial_{\gamma}B_{\gamma}X)=
-{4\gamma\over (1+\gamma^2)^2}$ because of (\ref{ScalarProduct}).
Therefore:
\begin{equation}
	\left.{\partial H_{\gamma}\over\partial \gamma}
	\right|_{X,\; \partial_{\tau}X}=
	{2\over 1+\gamma^2}{1\over 2\pi}
	\int d\sigma\left[\gamma(B_{\gamma}X,\partial_-X)
	+\gamma^{-1}(B_{\gamma}X,\partial_+X)\right]
\end{equation}
This is (\ref{Spectrality}), (\ref{LeftGenerating}).

\section{Conclusion.}
We have discussed the deviation of the classical string worldsheet
in AdS space from being periodic in the global time, which can be
quantitatively characterized as the action of the deck transformation
on the string phase space. 
This is important in AdS/CFT because it corresponds
to the anomalous dimension on the field theory side. 
We have considered the deck transformation on the {\em fast moving}
strings, in the null-surface perturbation theory.
We have shown that at least in the null-surface perturbation theory
the deck transformation can be understood
as an example of the B\"acklund transformation (and
in particular it can be connected to the identical transformation
by a continuous family of B\"acklund transformations).
This implies that the deck transformation  can
be understood as the finite time evolution by an infinite
linear combination of the local conserved charges. 

This can be easily understood in the plane wave limit, when
the string worldsheet theory becomes a collection of the 
free massive fields. In this case, the energy is given by
the formula:
\[
E=J+\sum_{I=1}^8 \sum_{n=-\infty}^{\infty}\overline{\alpha_n^I}\alpha_n^I
+\sum_{k=0}^{\infty} c_k{\cal G}_{2k}
\]
The first term on the right hand side is the $SO(2)$ angular momentum $J$
(integer), the second term is the total oscillator number (also integer),
and the third term, which measures the deviation of $E$ from
being integer, is an infinite sum of the local charges ${\cal G}_k$
with some coefficients $c_k$. 
On the field theory side the corresponding operator is of the form
\[
\mbox{tr}\;\cdots ZZZ\;\Phi\; ZZZ\cdots ZZ\; \Phi\; ZZZZ\cdots
\]
where $Z$ and $\Phi$ are complex scalars. From this point of view
$J$ is the total number of the letters $Z$ under the trace, and 
$\sum_{I=1}^8 \sum_{n=-\infty}^{\infty}\overline{\alpha_n^I}\alpha_n^I$
is the total number of the letters $\Phi$ (plus the total number
of derivatives, if we include derivatives acting on $Z$ and $\Phi$).
Therefore $\sum_{k=0}^{\infty} c_k{\cal G}_{2k}$ computes the anomalous 
dimension.

Having related the anomalous dimension to the B\"acklund transformations,
we study the basic properties of the B\"acklund transformations
and give an explicit formula (\ref{Spectrality}) for 
the ``generator'' of the B\"acklund transformation. 
Eq. (\ref{Spectrality}) is closely related to the general
property of B\"acklund transformations which is called
{\em spectrality} \cite{SklyaninBQ}. 
But in fact Eq. (\ref{Spectrality}) is somewhat different
from the usual definition of the spectrality.
The definition of \cite{SklyaninBQ} used 
the generating function which requires the separation of the
phase space variables into coordinates and momenta. 
We use instead the ``generator'' defined through 
the ``logarithm'' of the B\"acklund transformation.

The subject of B\"acklund transformations is one of the places where the
classical theory of integrable systems touches
the quantum theory.
Therefore we hope that the considerations presented in this paper
might be of some help in understanding the quantum worldsheet
theory for the superstring in $AdS_5\times S^5$.

\section*{Acknowledgments}
I want to thank N.~Beisert, C.~Nappi, A.~Rosly, 
A.~Tseytlin and K.~Zarembo for useful
discussions, and especially G.~Arutyunov for the correspondence
and for calling my attention to \cite{SklyaninBQ}. 
This research was supported by the Sherman Fairchild 
Fellowship and in part
by the RFBR Grant No.  03-02-17373 and in part by the 
Russian Grant for the support of the scientific schools
NSh-1999.2003.2.

%\bibliographystyle{utcaps}
%\bibliography{local}

\providecommand{\href}[2]{#2}\begingroup\raggedright\endgroup

\end{document}